\documentclass[a4paper,11pt]{article}
\usepackage{pos}
\usepackage{amsmath,mathtools,bbold}

\usepackage{float,wrapfig,subcaption}

\title{Logarithmic corrections to $a^2$ scaling in lattice QCD with Wilson and Ginsparg-Wilson quarks}
\ShortTitle{Logarithmic corrections to $a^2$ scaling in lattice QCD with Wilson and GW quarks}

\author*[a,b]{Nikolai Husung}
\author[a]{Peter Marquard}
\author[a,c]{Rainer Sommer}

\affiliation[a]{Deutsches Elektronen-Synchrotron DESY,\\
  Platanenallee 6, 15738 Zeuthen, Germany}

\affiliation[b]{Physics and Astronomy, University of Southampton,\\ Southampton SO17 1BJ, UK}

\affiliation[c]{Institut f\"ur Physik, Humboldt-Universit\"at zu Berlin,\\
 Newtonstr. 15, 12489 Berlin, Germany}

\emailAdd{n.husung@soton.ac.uk}
\emailAdd{rainer.sommer@desy.de}
\emailAdd{peter.marquard@desy.de}

\abstract{
We analyse the leading logarithmic corrections to the $a^2$ scaling of lattice artefacts in QCD, following the seminal work of Balog, Niedermayer and Weisz in the O(n) non-linear sigma model.
Limiting the discussion to contributions from the action, the leading logarithmic corrections can be determined by the anomalous dimensions of mass-dimension 6 operators.
These operators form a minimal on-shell basis of the Symanzik Effective Theory.
We present results for non-perturbatively O($a$) improved Wilson and Ginsparg-Wilson quarks.
\vspace{0.5cm}
\begin{flushright}
  DESY 21-178
\end{flushright}}

\FullConference{%
 The 38th International Symposium on Lattice Field Theory, LATTICE2021
  26th-30th July, 2021
  Zoom/Gather@Massachusetts Institute of Technology
}

\def\gbar{\bar{g}}
\def\ord{\mathrm{O}}
\def\MSbar{\ensuremath{\overline{\text{MS}}}}
\def\op{\mathcal{O}}
\def\opE{\mathcal{E}}
\def\base{\mathcal{B}}
\def\L{\mathscr{L}}
\def\Nf{N_\mathrm{f}}
\def\obs{\mathcal{P}}
\def\rmd{\mathrm{d}}
\def\tr{\mathop{\mathrm{tr}}}
\def\diag{\mathop{\mathrm{diag}}}

\begin{document}
\maketitle

\restylefloat{figure}

\section{Introduction}
\begin{wrapfigure}{r}{0.42\textwidth}
\includegraphics[scale=0.9,clip=true,trim=10 0 0 0]{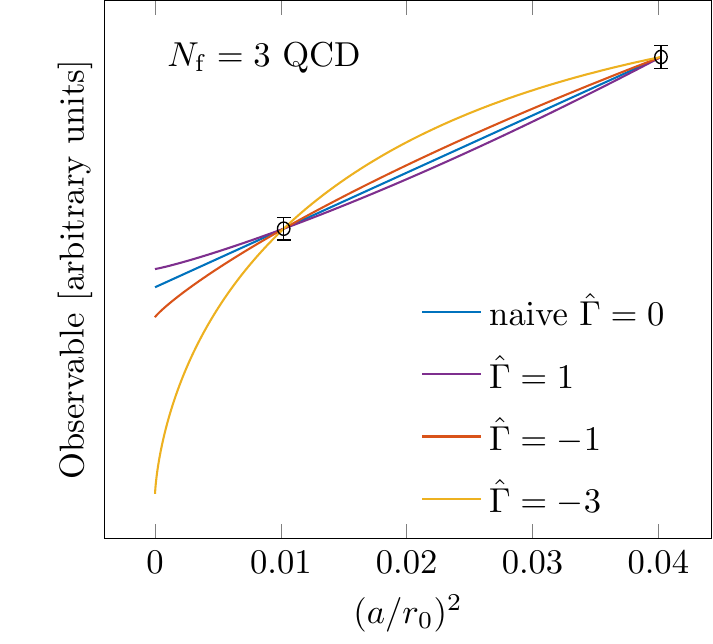}
\caption{Sketch of the asymptotic lattice spacing dependence $a^2[\gbar^2(1/a)]^{\hat{\Gamma}}$ for various values of $\hat{\Gamma}$ compared to the plain $a^2$.
We used here the 5-loop perturbative running of the QCD coupling in \MSbar{}~\cite{Luthe:2017ttc}.}
\label{fig:sketchImpact}
\end{wrapfigure}
Symanzik Effective Field Theory (SymEFT)~\cite{Symanzik:1979ph,Symanzik:1981hc,Symanzik:1983dc,Symanzik:1983gh} can be used to describe the lattice artifacts of lattice QCD for asymptotically small lattice spacings $a$.
In contrast to the (classical) $a^2$ ansatz commonly used in continuum extrapolations $a\searrow0$, the leading asymptotic lattice spacing dependence is actually of the form $a^2[\gbar^2(1/a)]^{\hat{\Gamma}}$ due to quantum corrections, where $\hat{\Gamma}$ is a real constant and $\gbar$ is the running coupling.
We assume here the use of fully $\ord(a)$ improved lattice actions throughout.
Knowing $\hat{\Gamma}$ is required to put continuum extrapolations on more solid grounds and to rule out any trouble arising from distinctly negative values for $\hat{\Gamma}$ as there is no theoretical lower bound on this value.
A particularly problematic example was found for the O(3) non-linear sigma model, where such an analysis~\cite{Balog:2009yj,Balog:2009np} was performed for the first time yielding $\hat{\Gamma}=-3$.
To highlight the impact a non-zero $\hat{\Gamma}$ can have, we added the oversimplified sketch in fig.~\ref{fig:sketchImpact} for the case of three-flavour QCD.

\section{Symanzik Effective Theory}
For a more complete picture of SymEFT see~\cite{Husung:2019ytz} as we give here only a short summary of the main concepts.
To describe the lattice artifacts we start from the effective Lagrangian
\begin{equation}
\L_\mathrm{Sym}=-\frac{1}{2g_0^2}\tr(F_{\mu\nu}F_{\mu\nu})+\bar{\Psi}\left\{\gamma_\mu D_\mu+m\right\}\Psi+a^2\sum_j c_j\op_j+\ord(a^3),
\end{equation}
which is just the (Euclidean) continuum QCD Lagrangian for $\Nf$ quark flavours $\Psi$ with additional $\ord(a^2)$ corrections.
The matching coefficients $c_j$ depend on the choice for the lattice discretisation.
(Only) For tree-level matching it suffices to naively expand the lattice action in the lattice spacing.
The basis of operators $\op_j$ must be chosen such that it parametrises all lattice artifacts originating from the lattice action up to higher order corrections in the lattice spacing.

Being interested in either Ginsparg-Wilson~(GW) or Wilson~\cite{Wilson:1974,Wilson:1975id} quarks for the lattice discretisation yields the following symmetry constraints on our minimal operator basis
\begin{itemize}
\itemsep0.5em
\item SU($N$) gauge symmetry,
\item invariance under Euclidean reflections,
\item invariance under charge conjugation,
\item H(4) lattice symmetry, i.e.~continuum O(4) symmetry is broken due to reduced rotational symmetry,
\item flavour symmetries, $\mathrm{SU}(\Nf)_\mathrm{L}\times\mathrm{SU}(\Nf)_\mathrm{R}\times\mathrm{U}(1)$ for massless GW quarks and $\mathrm{U}(\Nf)_\mathrm{V}$ for massless Wilson quarks.
\end{itemize}
Notice that $\mathrm{SU}(\Nf)_\mathrm{L}\times\mathrm{SU}(\Nf)_\mathrm{R}\times\mathrm{U}(1)\subset\mathrm{U}(\Nf)_\mathrm{V}$ such that the minimal basis of GW quarks is a subset of the full minimal basis needed for Wilson quarks.
Due to being only interested in on-shell physics we can make use of the continuum equations of motion to reduce the operator basis further~\cite{Luscher:1996sc}.

The minimal on-shell operator basis for the massless case (or sufficiently small quark masses) then is the following~\cite{Weisz:1982zw,Luscher:1984xn,Sheikholeslami:1985ij}
\begin{align}
\op_1&=\frac{1}{g_0^2}\tr(D_\mu F_{\nu\rho}D_\mu F_{\nu\rho}),&\op_2&=\frac{1}{g_0^2}\sum_\mu\tr(D_\mu F_{\mu\nu}D_\mu F_{\mu\nu}),\nonumber\\
\op_3&=\sum_\mu\bar{\Psi}\gamma_\mu D_\mu^3\Psi,&\op_{k\geq4}&=g_0^2(\bar{\Psi}\Gamma_k\Psi)^2,\label{eq:basis}
\end{align}
where $\Gamma_{4-7}\in\{\gamma_\mu,\gamma_5\gamma_\mu\}\otimes\{\mathbb{1},T^a\}$ and $\Gamma_{8-13}\in\{\mathbb{1},\gamma_5,\sigma_{\mu\nu}\}\otimes\{\mathbb{1},T^a\}$ with $\sigma_{\mu\nu}=\frac{i}{2}[\gamma_\mu,\gamma_\nu]$.
The operators $\op_2$ and $\op_3$ both break O(4) symmetry.
For massless GW quarks we only need $\op_{k\leq 7}$, while massless Wilson quarks require the entire set of operators listed here.
For the general massive case we get additional massive operators, that are listed and discussed in~\cite{H:inprep,S:inprep}.

\section{Leading powers in the coupling}\label{sec:LeadingPowers}
For an arbitrary Renormalisation Group invariant (RGI) spectral\footnote{For a non-spectral quantity also corrections from the local fields involved must be taken into account, which cancel out for spectral quantities.} quantity $\obs$ we may use the operator basis to write the leading lattice artifacts as
\begin{equation}
\obs(a)=\obs(0)-a^2\sum_j\bar{c}_j^\op\delta\obs_j^\op(1/a)\times\left\{1+\ord(\gbar^2(1/a))\right\}+\ord(a^3),
\end{equation}
where $\bar{c}_j^\op$ is the tree-level matching coefficient and $\delta\obs_j^\op$ contains the matrix elements of interest with an additional insertion of $\int\rmd^4x\,\op_j(x)$.
The remaining scale dependence of $\delta\obs_j^\op(1/a)$, where $1/a$ is the relevant renormalisation scale for lattice artifacts, is governed by the renormalisation group equation
\begin{equation}
\mu\frac{\delta\obs_i^\op(\mu)}{\rmd\mu}=-\left\{\gamma_0^\op\gbar^2(\mu)+\ord(\gbar^4)\right\}_{ij}\delta\obs_j^\op(\mu),
\end{equation}
where $\gamma_0^\op$ is the 1-loop coefficient of the anomalous dimension matrix.
In general $\gamma_0^\op$ is not diagonal, but in our case we can make a change of basis $\op\rightarrow\base$ such that $\gamma_0^\base=\diag\big((\gamma_0)_1,\ldots,(\gamma_0)_n\big)$ becomes diagonal.
In turn this allows to introduce the RGI, where all scale dependence is absorbed into some perturbatively known prefactor
\begin{equation}
\delta\obs_j^\base(1/a)=[2b_0\gbar^2(1/a)]^{\hat{\gamma}_j}\delta\obs_{j;\mathrm{RGI}}^\base\times\left\{1+\ord(\gbar^2(1/a))\right\},\quad \hat{\gamma}_j=\frac{(\gamma_0^\base)_j}{2b_0},
\end{equation}
where $b_0$ is the 1-loop coefficient of the $\beta$-function and the factor $2b_0$ in front of $\gbar^2(1/a)$ is the common choice for the normalisation.
Taking the leading order matching $c_j^\base(\gbar^2)=\hat{c}_j^\base[2b_0\gbar^2(1/a)]^{n_j^\mathrm{I}}\times\left\{1+\ord(\gbar^2(1/a))\right\}$ into account, we eventually arrive at the central formula for the leading asymptotic lattice spacing dependence
\begin{align}
\obs(a)&=\obs(0)-a^2\sum_j[2b_0\gbar^2(1/a)]^{\hat{\Gamma}_j}\hat{c}_j^\base\delta\obs_{j;\mathrm{RGI}}^\base\times\left\{1+\ord(\gbar^2(1/a))\right\}+\ord(a^3),\quad \hat{\Gamma}_j=\hat{\gamma}_j+n_j^\mathrm{I}\,,
\end{align}
which has precisely the form we mentioned in the beginning.
Of course there are now multiple $\hat{\gamma}_j$.
Those must be computed  to give a lower bound on these powers and to sort out, which one gives the leading contribution, if any $\hat{\gamma}_j$ is actually dominant.

\captionsetup{subrefformat=parens}
\begin{figure}
\begin{subfigure}[t]{0.19\textwidth}\centering
\includegraphics[scale=1]{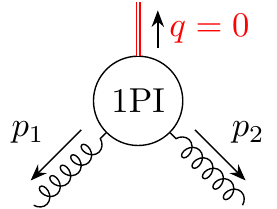}
\caption{}\label{fig:A2O}
\end{subfigure}
\begin{subfigure}[t]{0.19\textwidth}\centering
\includegraphics[scale=1]{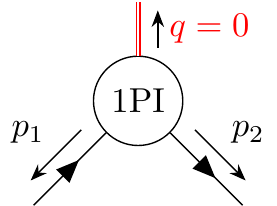}
\caption{}\label{fig:F2O}
\end{subfigure}
\begin{subfigure}[t]{0.19\textwidth}\centering
\includegraphics[scale=1]{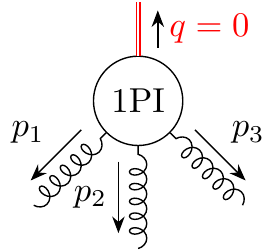}
\caption{}\label{fig:A3O}
\end{subfigure}
\begin{subfigure}[t]{0.19\textwidth}\centering
\includegraphics[scale=1]{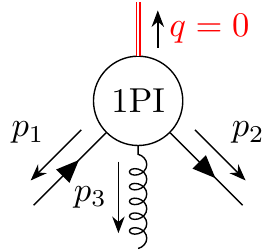}
\caption{}\label{fig:F2AO}
\end{subfigure}
\begin{subfigure}[t]{0.19\textwidth}\centering
\includegraphics[scale=1]{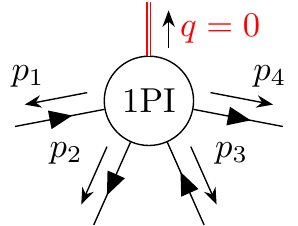}
\caption{}\label{fig:F4O}
\end{subfigure}
\caption{1PI graphs considered to perform the 1-loop renormalisation of the minimal operator basis at zero momentum.
The double line indicates the operator insertion at zero momentum.
Graph \subref{fig:F4O} is only needed to renormalise the 4-fermion opreators, while the graphs \subref{fig:A2O} and \subref{fig:A3O} would suffice for the case of pure gauge theory.}\label{fig:graphs}
\end{figure}

\subsection{Renormalisation strategy}
Our strategy to compute the 1-loop anomalous dimensions is based on the background field gauge~\cite{tHooft:1975uxh,Abbott:1980hw,Abbott:1981ke,Luscher:1995vs} in which we compute the one-particle-irreducible (1PI) graphs as depicted in fig.~\ref{fig:graphs}.
This particular choice allows us to easily perform the renormalisation of the inserted operator at zero momentum, which then allows us to ignore any mixing from total divergence operators.
Since we perform our operator renormalisation off-shell we have to take EOM vanishing operators $\opE$ into account, i.e.~the desired mixing matrix $Z^\op$ can be extracted from
\begin{equation}
\begin{pmatrix}
\op_i \\[6pt]
\opE_j
\end{pmatrix}_{\MSbar}=\begin{pmatrix}
Z_{ik}^\op & Z^{\op\opE}_{il} \\[6pt]
0 & Z^{\opE}_{jl}
\end{pmatrix}
\begin{pmatrix}
\op_k \\[6pt]
\opE_l
\end{pmatrix},
\end{equation}
where the subscript \MSbar{} indicates that we are using the \MSbar{} renormalisation scheme working in $D=4-2\epsilon$ dimensions.
The 1-loop coefficient of the anomalous dimension matrix can then be easily obtained from the mixing matrix
\begin{equation}
Z^\op=\mathbb{1}+\gamma_0^\op\frac{\gbar^2}{\epsilon}+\ord(\gbar^4).
\end{equation}

\subsection{Leading powers \boldmath$\hat{\Gamma}_j$}
Following the strategy described before, we are left with a range of values $\hat{\Gamma}_j$ and the (unknown) constants $d_j=\hat{c}_j^\base\delta\obs_{j;\mathrm{RGI}}^\base$.
If a matching coefficient $c_j^\base$ vanishes at tree-level, we assume the 1-loop order to be the first non-vanishing contribution -- of course those contributions could still be further suppressed.
For an in-depth discussion of $\hat{c}_j^\base$ for commonly used lattice discretisations, see~\cite{S:inprep}.
We will rather focus here on the spectrum $\hat{\Gamma}_j$ and try to make statements about the leading lattice artifacts ignoring potential hierarchies between different $\hat{c}_j^\base$.
The plots in figure~\ref{fig:spectra} show all powers $\hat{\Gamma}_j$ for $\ord(a)$ improved Wilson and GW quarks respectively up to N${}^3$LO contributions.
This is done to indicate the large spread of $\hat{\Gamma}_j$ at leading order, while anything beyond $\hat{\Gamma}_j\leq 1+\min_i\hat{\Gamma}_i$ will be hard to distinguish from e.g.~the NLO contributions of the truly leading powers.
Also the very dense spectrum at subleading orders becomes more apparent this way.

\def\scale{0.855}
\begin{figure}\centering
\includegraphics[scale=\scale,page=3]{addimages/MassDimension6heatmaps.pdf}\\[-6pt]
\includegraphics[scale=\scale,page=2]{addimages/MassDimension6heatmaps.pdf}
\includegraphics[scale=\scale,page=1,clip=true,trim=27 0 0 0]{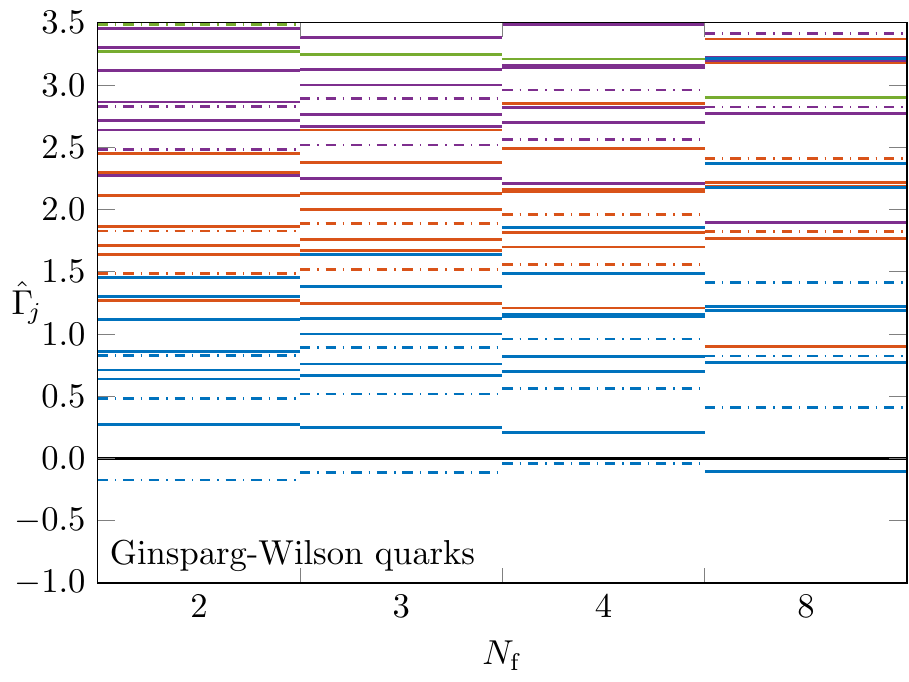}
\caption{Spectra of $\hat{\Gamma}_j$ for Wilson (left) and Ginsparg-Wilson quarks (right).
All powers up to N${}^3$LO have been plotted to highlight the spread of the leading powers and the density at subleading powers.
While the solid lines correspond to the contributions from the massless operator basis in eq.~\eqref{eq:basis}, the dash-dotted lines correspond to contributions from massive operators.
The number of flavours is chosen as $\Nf=2,3,4$ for the conventional lattice simulations and as $\Nf=8$ to highlight the approach to the conformal window.
Notice that due to the dense spectrum some contributions are hard to distinguish.}\label{fig:spectra}
\end{figure}

\section{Conclusion}
We find a very dense spectrum $\hat{\Gamma}_j$ for both Wilson and GW quarks due to the presence of four fermion operators at mass-dimension~6.
This will make it hard to decide, which contributions actually dominate the $\ord(a^2)$ lattice artifacts due to potentially complicated cancellations and pile-ups of the various contributions.
Nonetheless, ignoring any hierarchy between the matching coefficients, we find e.g.~for $\Nf=3$ the leading asymptotic dependence for spectral quantities (ordering $\hat{\Gamma}_i\leq\hat{\Gamma}_{i+1}$)
\begin{equation}
\frac{\obs(a)}{\obs(0)}=1-a^2[2b_0\gbar^2(1/a)]^{\hat{\Gamma}_\mathrm{min}}\left\{d_1+d_2 [2b_0\gbar^2(1/a)]^{\Delta\hat{\Gamma}}+\ldots\right\},\quad \begin{tabular}{lrr}
 & massless & massive \\
$\hat{\Gamma}_\mathrm{min}$ & $0.25$ & $-0.11$\\
$\Delta\hat{\Gamma}$ & $0.42$ & $0.36$
\end{tabular},
\end{equation}
which is universal for $\ord(a)$ improved Wilson and Ginsparg-Wilson quarks.
The asymptotic form for the massless case should also be a good approximation for $\Nf=2$ and may still work at $\Nf=2+1$ at physical quark masses.
Once the physical charm quark is added the contributions from massive operators will certainly not be small any longer and may actually be the dominant contributions.

For the massless case and $\Nf=2,3,4$ the convergence towards the continuum limit should be slightly improved due to $\hat{\Gamma}_i>0$, while both $\Nf=8$ and the massive case have slightly negative $\hat{\Gamma}_i\gtrsim -0.2$, such that the convergence might be worse.
In contrast to the O(3) non-linear sigma model~\cite{Balog:2009yj,Balog:2009np} all leading powers are very close to the classical zero and not distinctly negative, i.e.~$\hat{\Gamma}_i\gg-3$, which is good news.

When the different constants $d_j$ have a similar magnitude, the leading power in the coupling dominates the $a^2$ effects.
However, as analysed in some detail in~\cite{S:inprep}, common lattice actions can have $\hat{c}_j^\base$ which differ very much.
For example for an $\ord(a)$ improved fermion action and an improved gauge action, a single term dominates and it does not have the leading power.
Such information should be incorporated when continuum extrapolations are performed and checks on contaminations
of $\ord(a^3)$ or $\ord(a^4)$ contributions are advisable as well.
Necessary extensions to this work are amongst others the inclusion of contributions from local fields to go beyond spectral quantities and staggered quarks, which require an enlarged operator basis due to flavour changing interactions.

\textbf{Acknowledgements:} We thank Hubert Simma, Kay Sch\"onwald and Agostino Patella for useful discussions and suggestions.
RS acknowledges funding by the H2020 program in the  {\em Europlex} training
 network, grant agreement No. 813942.

\bibliographystyle{JHEP}
\bibliography{lat21-husung.bbl}

\end{document}